\documentclass[%
 reprint,
 amsmath,amssymb,
 aps,
]{revtex4-2}
\long\def\comment#1{}

\usepackage{graphicx}
\usepackage[latin1]{inputenc}
\usepackage{amsfonts}
\usepackage{amsmath,amssymb}
\usepackage{bm}
\usepackage{ulem}
\usepackage{color}
 \long\def\comment#1{}

\begin{document}
\graphicspath{{images/}}

\title{Charging capacitors from thermal fluctuations using diodes}
\author{P.M. Thibado,$^{1,*}$ J. C. Neu,$^2$ Pradeep Kumar,$^1$ Surendra Singh,$^1$ and L. L. Bonilla$^3$}
\affiliation{$^1$Department of Physics, University of Arkansas, Fayetteville, Arkansas 72701, USA.\\
$^2$Department of Mathematics, University of California, Berkeley, California 94720, USA.\\
$^3$G. Mill\'an Institute for Fluid Dynamics, Nanoscience and Industrial Mathematics and Department of Mathematics, Universidad Carlos III de Madrid, 28911 Legan\'es, Spain.\\
$^*$Corresponding author. E-mail: thibado@uark.edu}
\date{\today}

\begin{abstract}
We theoretically consider a graphene ripple as a Brownian particle coupled to an energy storage circuit. When circuit and particle are at the same temperature, the second law forbids harvesting energy from the thermal motion of the Brownian particle, even if the circuit contains a rectifying diode. However, when the circuit contains a junction followed by two diodes wired in opposition,  the  approach to equilibrium may become ultraslow. Detailed balance is temporarily broken as current flows between the two diodes and charges storage capacitors. The energy harvested by each capacitor comes from the thermal bath of the diodes while the system obeys the first and second laws of thermodynamics.
\end{abstract}

\maketitle

Numerous sources of ambient energy including kinetic, solar, ambient radiation, acoustic, thermal, etc. are readily available for energy harvesting. Energy harvesting in a quiet, dark setting is the most challenging because only thermal energy is present. In such an environment the Brownian motion of electrons produces a stochastic alternating current \cite{johnson,nyquist}. If this signal is rectified, energy could be harvested by charging a capacitor.
Using a diode to rectify noise in thermal equilibrium was ruled out by Brillouin because it violates detailed balance \cite{brillouin}. Gunn  added more insight by showing that diode nonlinearity generates an oppositely flowing current that cancels out the conventional rectified current \cite{gunn68, gunn69}. Feynman popularized the notion that it's impossible to harvest thermal energy at a single temperature in his lecture series ``Ratchet and pawl" \cite{feynman}. 

Renewed interest in thermal energy harvesting emerged in the 1990s, when it was discovered that diodes can rectify stochastic signals provided long-time correlations (non-white noise) are present \cite{magnasco, doe94}. More recently, it was discovered that electrical circuits containing multiple loops can give rise to unusual correlations with vortex dynamics~\cite{fil07,chi17,gon19}. This fueled further interest in this problem.

The simplest nonlinear circuit that can potentially store charge has a diode and a capacitor. The master equation for this circuit was first derived and studied by van Kampen in 1960~\cite{kam60}. He showed in equilibrium the capacitor has zero charge, and developed an approximate Fokker-Planck equation (FPE) that does not satisfy the fluctuation-dissipation theorem. Later a diode-capacitor-resistor circuit was studied by Sokolov in the late 1990s~\cite{sok98,sok99}. He derived a FPE that satisfies the fluctuation-dissipation theorem. In his study, the resistor and diode are held at different temperatures and the steady-state heat engine efficiency is determined. What has not been studied thus far is the full transient response of the charge on a storage capacitor for various diode-capacitor systems held at a single temperature.

In this study, we present a system capable of harvesting energy from thermal noise at a single temperature without violating the first or second law. Our system uses a small variable capacitor wired to two diodes and two storage capacitors using two current loops. Surprisingly, the nonlinearity of the diodes combined with the multiple current paths charges the capacitors with an ultraslow convergence to equilibrium. The harvested energy comes from the thermal baths of the diodes \cite{sekimoto}.

\begin{figure}[ht]
\begin{center}
\includegraphics[width=8cm]{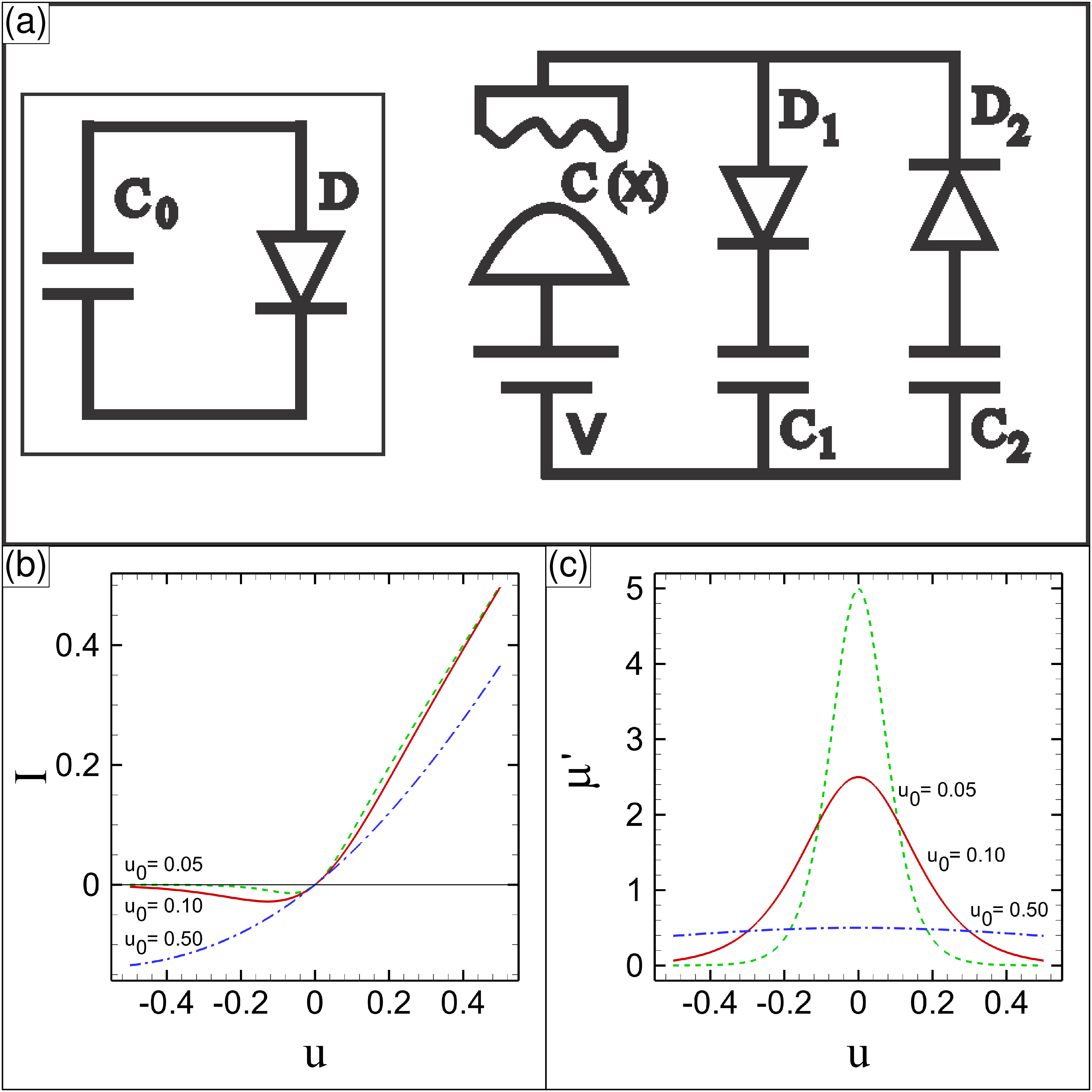}
\end{center}
\caption{Circuits and diode characteristics using $R = 1$. (a)~Two-diode circuit with variable capacitance graphene membrane. The inset is a single diode with a capacitor.  (b)~Current-voltage characteristics of the diodes used in this study for three values of diode parameter $u_0$. (c)~Plots of the derivative of the diode conductance $\mu'$ are shown for the same three values of $u_0$. \label{fig1}}
\end{figure}

The two current loop circuit model used for this study is shown in Fig.~\ref{fig1}(a). It includes a capacitor formed by a graphene membrane suspended near a STM tip and a DC bias voltage ($V$), which can be used to alter the average charge on the graphene capacitor. Due to thermal fluctuations, the distance between the graphene membrane and electrode changes, giving rise to a variable capacitance that can be written as  $C(x)=C_0/(1+\frac{x}{d})$, where $C_0=\varepsilon A/d$, $\varepsilon$ is the permittivity, $A$ is the effective area,  $d$ is the fixed distance between the membrane support and the tip,  and $x(t)$ is the graphene position. The series combination of $V$ and $C(x)$ acts as an  AC power source, as charge must flow on and off the capacitor according to $q(t)=C(t)V$ \cite{phi77, har20}. Earlier experimental and theoretical studies of this circuit using scanning tunneling microscopy and Langevin equation found that graphene's movement redistributes its thermal power to technologically important lower frequencies \cite{thi20, ack16}. This, coupled with graphene's unusual flexibility, makes it an ideal kinetic source of energy.

We have altered the earlier circuit  \cite{thi20} to include storage capacitors for energy harvesting. To achieve this, it is crucial to induce long lived transients during which the capacitors charge before they discharge to reach overall thermal equilibrium.  
The key to this is to connect the diodes to capacitors in the circuit  such that  the current passes only in one direction.
This allows transient charging of capacitors. Of course, real diodes leak in the low conducting region, and the system will eventually relax to thermal equilibrium. The circuit presented here maximizes the transient stage where the capacitors can charge and store energy amenable to harvesting.

At the graphene-diode junction in Fig.~\ref{fig1}(a), current can either flow through diode $D_1$ and charge storage capacitor $C_1$,  or flow through $D_2$ and charge $C_2$. The diodes are wired in opposition. Forward current senses minimal resistance when flowing either from $D_1$ to $C_1$ or from $C_2$ to $D_2$.  This circuit with diodes, storage capacitors  and variable  graphene capacitor is potentially a full-wave rectifying energy harvesting circuit. We track the harvested energy of the circuit using the Hamiltonian:
\begin{eqnarray}
\mathcal{H}_q(q,q_1,q_2)=\frac{q^2}{2C(x)}+\frac{q_1^2}{2C_1}+\frac{q_2^2}{2C_2}+qV, \label{eq1}
\end{eqnarray}
where $q$, $q_1$, and $q_2$ are the charges on the graphene and the two storage capacitors. The performance of the circuit depends on the current-voltage characteristics of the diodes. Each diode current $I_i$ is related to diode conductance $\mu_i$  and voltage drop across the diode $u_i$ via 

\begin{eqnarray}
&& I_1(u_1)=\mu_1(u_1) u_1, \; I_2(u_2)=\mu_2(-u_2) u_2, \quad \label{eq2} \\
&&\mu(u) = \frac{1}{R} \frac{1}{1+e^{-u/u_0}},  \label{eq3}
\end{eqnarray}
where the minus sign in diode two aligns the forward bias direction to be opposite diode one, and the diode conductance $\mu(u)$ is modeled as a sigmoid with parameter $u_0$, which controls how leaky the diodes are in reverse bias, as shown in the $I-u$ curves in Fig.~\ref{fig1}(b). The current-voltage curve is similar to an ideal diode in series with a resistor, which is more realistic, and except for the resistance $R$ it depends on a single parameter $u_0$ \cite{sze81}. It is not necessary to use the sigmoid function, and capacitor charging is found using only a polynomial expansion of the ideal diode formula. Also shown in Fig.~\ref{fig1}(c) is the derivative of the diode conductance $\mu'=d\mu/du$, which is a key function in the energy harvesting process. Note that while the current at zero volts is zero [Fig.~\ref{fig1}(b)], the derivative of the diode conductance $\mu'$ is nonzero at zero volts and its value increases as $u_0$ decreases.

The diode voltages follow from Kirchhoff's loop law (or derivatives of the Hamiltonian with respect to charges $q_1$ and $q_2$): 
\begin{eqnarray}
u_i =-\frac{\partial\mathcal{H}_q}{\partial q_i}= -\!\left(\frac{q_i}{C_i}+V+\frac{q_1+q_2}{C(x)}\right)\,, \label{eq4}
\end{eqnarray}
where we have used $q = q_1 + q_2$, from Kirchhoff's junction law $I=I_1+I_2$. 

The probability density $\rho(q_1,q_2,t)$  for capacitor charges $q_1$ and $q_2$  obeys  the Fokker-Planck equation (FPE) derived from an electron master equation in the continuum limit \cite{thi20}:
\begin{eqnarray}
\frac{\partial \rho}{\partial t}+\frac{\partial j_1}{\partial q_1} +\frac{\partial j_2}{\partial q_2} =0\,,
\label{eq5}
\end{eqnarray}
where $j_i=\mu_i(u_i)(u_i \rho-k_BT\partial_i \rho)$. See Appendix \ref{ap_a} for the full system FPE. The relaxation time of graphene is much smaller than the circuit $RC$ time; therefore, it reaches equilibrium much faster. In this limit, stochastic averaging~\cite{kha64,bon14} over the graphene dynamics allows us to replace $C(x)$ with $C_0$ provided the variation of $x(t)$ is small compared to $d$; see Appendix \ref{ap_a}. This is confirmed by numerical simulation of the Ito SDEs associated with the FPE; see Appendix \ref{ap_f}.  We simplify this presentation further and set $V=0$ to study energy harvesting solely from the thermal environment. We also studied the role of adding a nonzero bias voltage. The main outcome is to alter the final equilibrium charge on the storage capacitor to be $q = C_0 V$, instead of $q = 0$.

\begin{figure}[ht]
\begin{center}
\includegraphics[width=8cm]{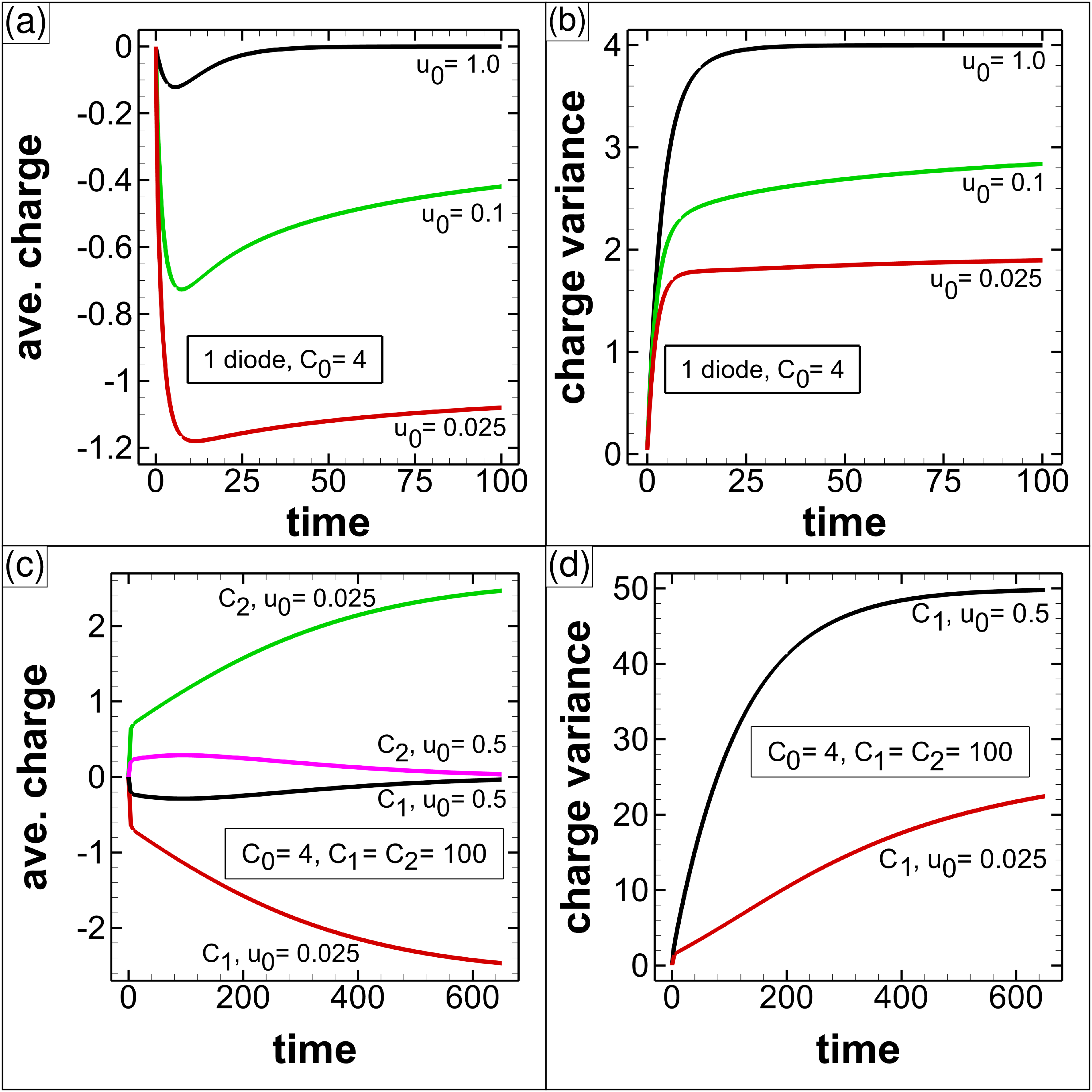}
\end{center}
\caption{Numerical solution of Fokker-Planck equation using $k_BT=1$, $R=1$, and various diode parameter values are displayed. (a) The average charge on the capacitor and (b) charge variance as functions of time for the single diode-capacitor circuit with different values of $u_o$ as labeled. (c)~The average charges on the capacitors and (d) their variance for the full two-diode circuit as functions of time for two different values of $u_0$ as labeled. \label{fig2}}
\end{figure}

To highlight the special features of the two-loop circuit, we first present numerical solutions for the one diode one capacitor circuit (inset of Fig.~\ref{fig1}(a)). The FPE for this circuit is given by~(see Appendix \ref{ap_b})
\begin{equation}
\frac{\partial \rho}{\partial t}=\frac{\partial }{\partial q}\!\left[\mu_0\!\left(\frac{q}{C_0}\rho+k_BT\frac{\partial  \rho}{\partial q}\right)\right]\!,\, \mu_0=\mu (-q/C_0)\,.\label{eq6}
\end{equation}
The average charge on the graphene capacitor in time is shown in Fig.~\ref{fig2}(a) for three different $u_0$ values for fixed $C_0=4$, $k_BT=1$, and $R=1$.  We have chosen parameters that allow the simulation to capture the important physics in a qualitative manner. A quantitative comparison is made later. The unit of charge is given by $\sqrt{k_BTC_0}$ and the unit of time is given by $RC_0$. In all three cases, the average charge on the capacitor increases from zero to a negative maximum, then decays to equilibrium value zero. The relative sign of the charge reflects the choice of the positive direction for the current.  The smallest value of the diode parameter ($u_0=0.025$) achieves the largest charge before slowly relaxing to zero. For the largest value of $u_0$, the charge reaches the equilibrium in the shortest time. 

The variance of the capacitor charge for the same diode parameters is shown in Fig.~\ref{fig2}(b). It grows monotonically in time.  For a perfect diode, the circuit would reach thermal equilibrium for $\left<q\right><0$.  This corresponds to probability density $\rho_\infty=e^{-q^2/(2C_0k_BT)}\Theta(-q)\sqrt{\pi C_0k_BT/2}$, where $\Theta(x)$ is the Heaviside unit step function. The average charge of this distribution is $\langle q\rangle=-\sqrt{2k_BTC_0/\pi}$, and the variance, $\langle (q- \langle q \rangle)^2\rangle=k_BTC_0(1-2/\pi)$. The maximum average charge and variance in Fig.~\ref{fig2}(a) and \ref{fig2}(b) tend to these values as $u_0\to 0$. In the small temperature limit, $k_BT\ll C_0u_0^2$, it is possible to describe analytically the evolution of the probability density from an initial condition; see Appendix \ref{ap_b}. It is a front that leaves the equilibrium density at its rear end. Its forefront is a Gaussian that advances slowly and narrows as it propagates.

For the full circuit with two diodes and three capacitors, a new charging dynamic arises, as shown in Fig.~\ref{fig2}(c). The graphene capacitance is kept at $C_0=4$, while the storage capacitances are set at $C_1=C_2=100$.  The average charges on $C_1$ and $C_2$ as functions of time are shown in Fig.~\ref{fig2}(c) for two different diode parameters. In the initial charging phase, we see a rapid increase in charge. Charges  $q_1$ (negative) and $q_2$ (positive) are perfectly anti-correlated with each capacitor storing an equal amount of energy. However, after the initial rapid charging phase the charge does not monotonically decay to zero as in the one-diode-capacitor circuit. Depending on the value of $u_0$, it may continue to increase before reversing to relax to zero. To illustrate dependence on the diode parameter $u_0$, a second set of charging curves is shown in Fig.~\ref{fig2}(c) with a larger $u_0$. In this case, the charge decays more quickly to zero. The variance of the charge in time for both values of $u_0$  increases monotonically, as shown in Fig.~\ref{fig2}(d).  The variance  eventually reaches the same equilibrium value for both.

To understand the origin of the initial rapid rise of the capacitor charge, we consider the equation for the average charge on the storage capacitor, which follows from  Eq. (5), %
\begin{eqnarray}
\frac{d}{dt}\langle q_i \rangle = \langle u_i \mu_i(u_i)\rangle - k_B T \left(\frac{1}{C_0} + \frac{1}{C_i}\right) \langle\mu'_i(u_i) \rangle{\color{blue}.} \label{eq7}
\end{eqnarray}
The first term  on the right is the conventional Ohm's law current. The second term is the nonlinear thermal current proportional to the temperature. From this equation, the initial charge growth, with $\rho = \delta(q_1) \delta(q_2)$  is given by  $\left. \frac{d}{dt}\langle q_i \rangle\right|_{t=0} = - k_B T (\frac{1}{C_0} + \frac{1}{C_i}) \mu'_i(0)$.  This shows that the initial charging of the capacitors  is possible only for nonlinear resistive devices with nonzero conductance slope $\mu'_i(0)=(-1)^{i+1}/(4Ru_0)$, which from  Fig.~\ref{fig1}(c) can be sizable for small $u_0$.  The initial diode current flows opposite to the conventional current and puts negative charges  on storage capacitor $C_1$. A detailed discussion of initial charging for both one diode and one capacitor as well as the three capacitor systems is provided in~Appendix \ref{sec_c} and \ref{ap_d}.

\begin{figure}[ht]
\begin{center}
\includegraphics[width=8cm]{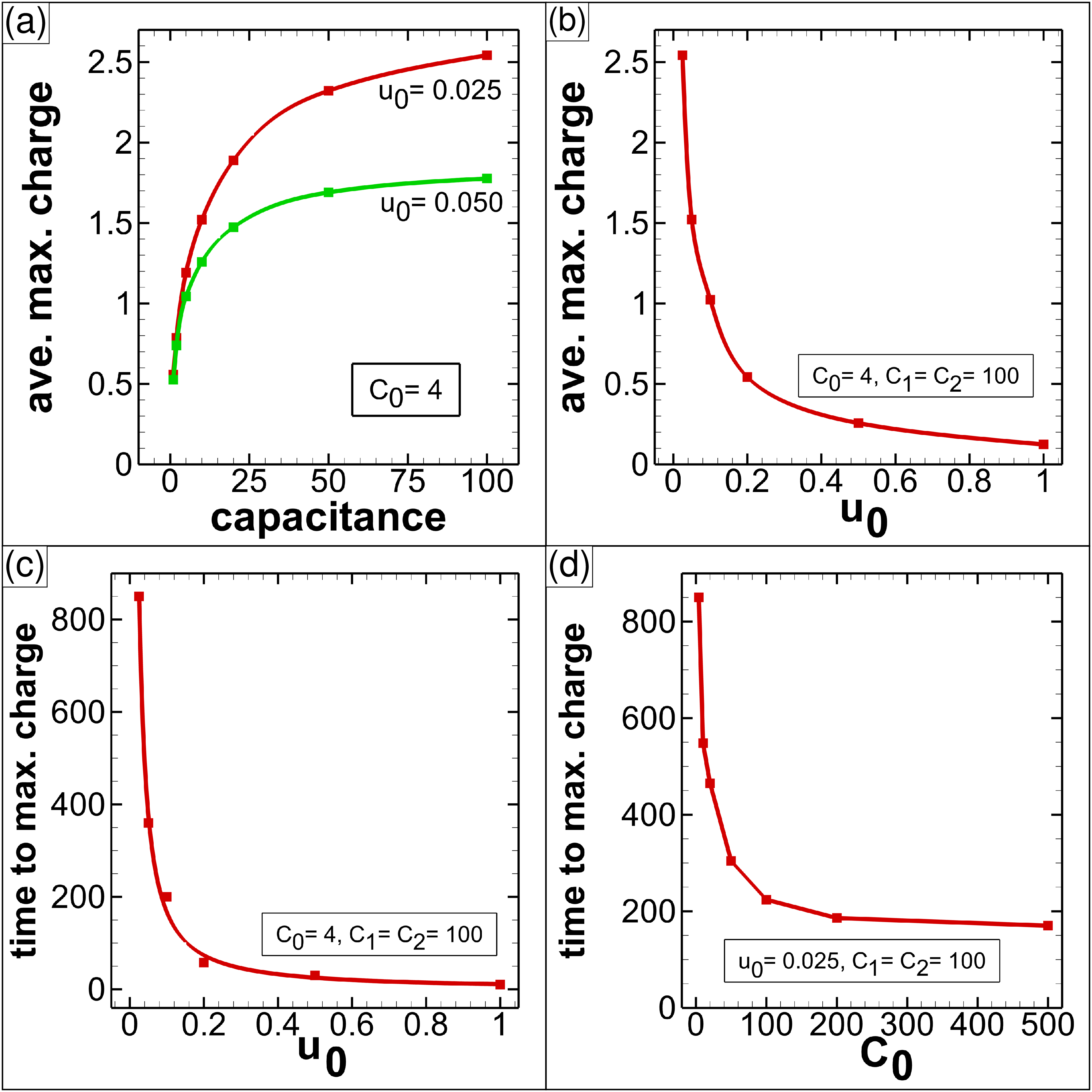}
\end{center}
\caption{Numerical solution of Fokker-Planck equation for our full circuit model using $k_BT=1$ and $R=1$. (a) Maximum average charge on the storage capacitors as a function of capacitance for two different values of the diode parameter $u_0$. (b) Dependence of maximum average charge on storage capacitors on diode parameter $u_0$. (c) Time to reach the maximum charge as a function of diode parameter $u_0$. (d) The time to reach the maximum charge vs graphene capacitance parameter $C_0$.}
\label{fig3}
\end{figure}
\begin{figure}[ht]
\begin{center}
\includegraphics[width=7cm]{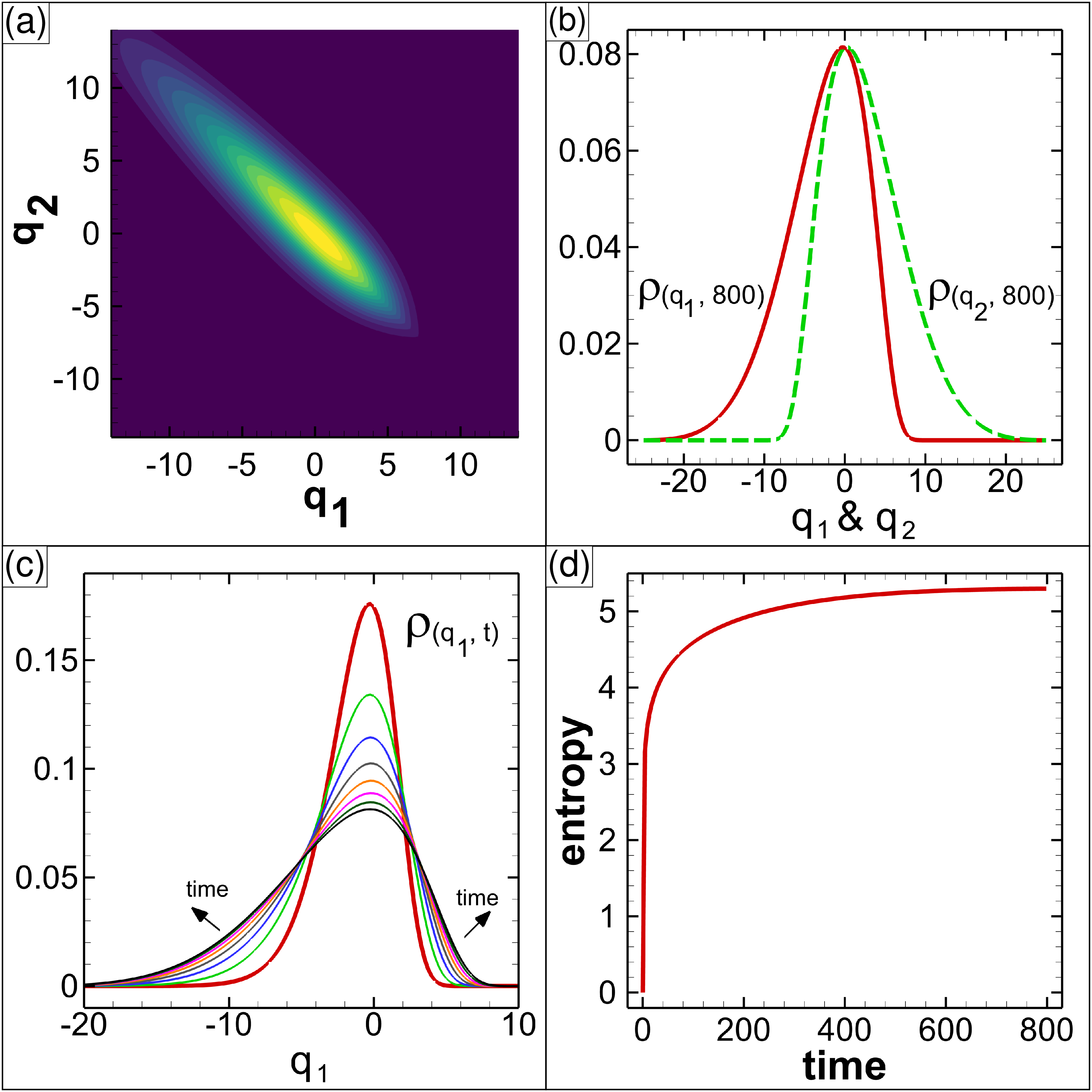}
\end{center}
\caption{Numerical solution of Fokker-Planck equation for our full circuit model using $k_BT=1$ and $R=1$. (a) Two-dimensional charge distribution $\rho(q_1,q_2,800)$. (b) Charge distributions $\rho(q_1,800)$ (full curve) and $\rho(q_2,800)$ (dashed curve) for the storage capacitors at the time of maximum charge. (c)~Time evolution of the charge distribution for the storage capacitors shown in equal time intervals from 100 to 800. (d)~Time evolution of  Shannon entropy. }    
\label{fig4}
\end{figure}

The dependence of maximum charge on various circuit parameters is explored in Fig.~\ref{fig3}. The  maximum average charge increases with the charging capacitance, as shown in Fig.~\ref{fig3}(a) for two different diode parameters. The maximum charge as a function of the diode parameter for a fixed storage capacitance decreases as shown in Fig.~\ref{fig3}(b). The time to reach maximum charge increases as $u_0$ decreases, as shown in Fig.~\ref{fig3}(c). The time to reach maximum charge is very sensitive to the value of $C_0$. Here, the smaller the value of $C_0$, the longer the capacitors remain charged, which is opposite to the the single diode case. In summary, better performance of the circuit in regard to energy harvesting is achieved for small $u_0$ and small values of the ratio $C_0/C_1$.
To understand the mechanism behind the charging of the storage capacitors, time evolution of the probability distributions of charges must be considered. For the plot of Fig.~\ref{fig2}(c) with $u_0=0.025$, we present various plots of the probability density of charges. Fig.~\ref{fig4}(a) shows the two dimensional probability density, $\rho(q_1,q_2,t)$ for $t=800$. The probability density is symmetric about $q_1=-q_2$. The one dimensional marginal charge distributions for each storage capacitor $\rho(q_1,800)$ and $\rho(q_2,800)$ is shown in Fig.~\ref{fig4}(b). In the limit as the diode parameter $u_0$ is reduced to zero and the time increased, these distributions approach one sided Gaussian distributions.  The two mirror one another.

Fig.~\ref{fig4}(c) shows the marginal charge distribution $\rho(q_1,t)$ every $100$ time units. The distribution spreads out toward the thermal equilibrium distribution; however, it is apparent that the spread slows down. Just as the variance takes an extremely long time to reach equilibrium value (Fig.~\ref{fig2}(d)), the right side of the distribution in Fig.~\ref{fig4}(c) slows down. Thus, the two-diode circuit has created an ultraslow approach to equilibrium. From the time dependent probability distributions, we can calculate the evolution of the Shannon entropy, which is shown in Fig.~\ref{fig4}(d)~(see Appendix \ref{ap_d})~\cite{seif12}. The entropy monotonically increases in agreement with the second law, and approaches the equilibrium value in the long time limit. 

Energy transferred to the storage capacitors came from the thermal bath. For the single ideal-diode-capacitor case, the energy harvested can be found analytically to be $k_BT/\pi$. More importantly, the power is found to be $k_BT/(\pi RC)$. A series of these units may be built on an integrated circuit with each using a space of less than 0.1 square microns, with the potential to produce a significant power density~(see Appendix \ref{ap_f}) \cite{fer20}. Surprisingly, a recent study found the average power density for wind and solar farms is relatively low, at 0.50 and 5.4 W/m$^2$ \cite{mil18}.

The FPE \eqref{eq5} has a unique stable equilibrium solution corresponding to the minimum free energy and zero average stored charge~(see Appendix \ref{ap_d}). However, before reaching thermal equilibrium, the circuit of Fig.~\ref{fig1}(a) for a small value of $u_0$ produces a long-lived transient state in which the capacitors store charges proportional to their capacitance. Pushing off equilibrium for a sufficient time period allows the storage capacitors to be disconnected from the circuit and the energy taken from the thermal surroundings to be used. For energy harvesting, the circuit topology investigated here enhances the total charge and time to charge, while providing additional handles for manipulating the outcome. It would be interesting to study other circuit designs, such as multiple stages of our circuit connected together in parallel or series. It is feasible to build  a circuit array  at a foundry as an integrated circuit using silicon fixed capacitors. This study demonstrates that energy can be harvested from the thermal surroundings at a single temperature without violating the laws of thermodynamics.



In an earlier study~\cite{har20}, we used a similar circuit to charge capacitors using a variable capacitor driven by a motor. The source of power charging the storage capacitors was the motor. In contrast, here for the first time, we demonstrate that the ambient thermal environment can be a source of power for charging storage capacitors. We hold the circuit and environment at the same temperature and prove that charging the capacitors does not violate the laws of thermodynamics. Furthermore, we have discovered a circuit topology that provides additional degrees of freedom which enable ultrafast charging of the storage capacitors combined with ultraslow convergence to equilibrium. These aspects have practical significance, as they allow time to disconnect the storage capacitors from the circuit for possible energy harvesting before they lose their charge.

In summary, we have studied theoretically the spontaneous thermal fluctuations of a circuit with diodes having nonlinear current-voltage characteristics and storage capacitors. If the storage capacitors have an initial charge of zero, the circuit draws power from the thermal bath to charge them. Throughout the process, the system satisfies both the first and second laws of thermodynamics. From Brillouin, as mentioned earlier, we know that the diode nonlinearity generates an oppositely flowing current which exactly cancels out the rectified current to maintain detailed balance. However, this opposite current also initially charges the storage capacitors. The larger the storage capacitance, the more charge it can harvest. A smaller graphene capacitance provides a higher initial rate of charging. In addition, a smaller graphene capacitance yields a longer charging time.

\acknowledgments
This project was supported by the Walton Family Charitable Support Foundation. We acknowledge support by the FEDER/Ministerio de Ciencia, Innovaci\'on y Universidades--Agencia Estatal de Investigaci\'on Grant No. PID2020-112796RB-C22, by the Madrid Government (Comunidad de Madrid-Spain) under the Multiannual Agreement with UC3M in the line of Excellence of University Professors (EPUC3M23), and in the context of the V PRICIT (Regional Programme of Research and Technological Innovation). The authors thank the Arkansas high performance computing center.

\appendix
\section{Eliminating graphene local equilibrium from circuit}\label{ap_a}
The Fokker-Planck equation (FPE) of the full system comprising graphene and circuit  in Fig.~\ref{fig1} is \cite{thi20}
\begin{widetext}
\begin{subequations}\label{eqa1}
\begin{eqnarray}
&&\frac{\partial\rho}{\partial t}-\sum_{i=1}^2\!\frac{\partial}{\partial q_i}\!\left[\mu_i(u_i)\left(\rho \frac{\partial\mathcal{H}}{\partial q_i}+k_BT \frac{\partial\rho}{\partial q_i}\right)\right]+\frac{p}{m}\frac{\partial\rho}{\partial x} -\frac{\partial}{\partial p} \left[\rho\frac{\partial\mathcal{H}}{\partial x}+\eta\left(\frac{p}{m}\rho+k_BT\frac{\partial\rho}{\partial p}\right)\! \right]=0,\quad \label{eqa1a}\\
&&\mathcal{H}=\frac{p^2}{2m}+ U(x)-\frac{C_0V^2x}{2d}+\frac{(q_1+q_2)^2}{2C(x)}+\sum_{j=1}^2\frac{q_j^2}{2c_j}+(q_1+q_2)V.        \label{eqa1b}
\end{eqnarray}
That the equilibrium $\rho_\text{eq}\propto e^{-\frac{\mathcal{H}}{k_BT}}$ is a globally stable solution of Eq.~\eqref{eqa1a} can be proved by showing that the relative entropy,
\begin{equation}
\mathcal{F}[\rho](t)=-k_B\int \rho(x,p,q_1,q_2,t)\, \ln\!\left(\frac{\rho(x,p,q_1,q_2,t)}{\rho_\text{eq}(x,p,q_1,q_2)}\right)dx\, dp\, dq_1dq_2, \label{eqa1c}
\end{equation}
is a Lyapunov functional of Eq.~\eqref{eqa1a}; see \cite{ris84}.
\end{subequations}

\begin{table}[ht]
\begin{center}\begin{tabular}{cccccc}
 \hline
$x$&  $p$ & $q, q_i$ &$t$& $\mathcal{H}$ & $V$\\ 
$l$ & $\sqrt{mT}$& $C_0V_0$  &$RC$ & $k_BT$ & $V_0$  \\ 
\hline
\end{tabular}
\end{center}
\caption{Units for nondimensionalizing the equations of the model.  $C_0V_0^2=k_BT$, $R=1/\mu_f$.}
\label{table1}
\end{table}
Using the nondimensional units of Table \ref{table1} in Eqs.~\eqref{eqa1}, we obtain 
\begin{subequations}\label{eqa2}
\begin{eqnarray}
&&\varepsilon\left\{\frac{\partial\rho}{\partial t}-\nu\sum_{i=1}^2\!\frac{\partial}{\partial q_i}\!\left[\mu_i\left(\rho \frac{\partial\mathcal{H}}{\partial q_i}+ \frac{\partial\rho}{\partial q_i}\right)\right]\right\} +\delta p\frac{\partial\rho}{\partial x} -\frac{\partial}{\partial p} \left[\delta\,\rho\frac{\partial\mathcal{H}}{\partial x}+p\rho+\frac{\partial\rho}{\partial p} \right]=0,\quad \label{eqa2a}\\
&&\mathcal{H}=\frac{p^2}{2}+ U(x)-\frac{x}{2\lambda}+\frac{(q_1+q_2)^2}{2\Gamma(v)}\left(1+\frac{x}{\lambda}\right)+\nu \sum_{j=1}^2q_j^2+(q_1+q_2)v,       \label{eqa2b}\\
&&\varepsilon=\frac{m}{\eta RC},\quad \delta= \frac{U_B\sqrt{m}}{\eta l\sqrt{k_BT}},\quad\lambda=\frac{d}{l},\quad\nu=\frac{C_0}{C},\quad v=\frac{V}{V_0}, \quad V_0=\sqrt{\frac{k_BT}{C_0}}.     \label{eqa2c}
\end{eqnarray}
\end{subequations}\end{widetext}

We now derive averaging formulas for the FPE and the Ito stochastic differential equations (SDEs) \cite{gardiner} assuming $\varepsilon\ll 1$.  This is reasonable as $m/\eta$ is the reciprocal of the phonon frequency (picosecond scale) and $RC$ is typically on the nanosecond scale. In a fast time scale $\tau=t/\varepsilon$, the leading order probability density evolves to the local equilibrium $\Upsilon$ below. Inserting
\begin{subequations}\label{eqa3}
\begin{eqnarray}
&&\rho=\Upsilon(x,p,q_1+q_2)\tilde{\rho}(q_1,q_2,t) +\varepsilon f(x,p,\mathbf{q}), \label{eqa3a}\\
&&\Upsilon(x,p,q)=\frac{1}{Z(q)}e^{-\mathcal{H}_r}, \label{eqa3b}\\
&&\mathcal{H}_r=\frac{p^2}{2}+ U(x)-\frac{x}{2\lambda}+ \frac{q^2}{2\Gamma(v)}\!\left(1+\frac{x}{\lambda}\right)\!, \label{eqa3c}\\
&&\int\Upsilon(x,p,q)\, dx\, dp=1,\quad\int \tilde{\rho}(q_1,q_2)\, dq_1dq_2=1,\quad \label{eqa3d}
\end{eqnarray}\end{subequations}
into Eq.~\eqref{eqa1a}, we obtain 
\begin{eqnarray}
&& \frac{\partial}{\partial p} \left(f\frac{\partial\mathcal{H}}{\partial x}\delta+ pf+\frac{\partial f}{\partial p}\right)\! -p\frac{\partial f}{\partial x}\delta\nonumber\\
&&\quad= \Upsilon\,\frac{\partial\tilde{\rho}}{\partial t}-\nu\sum_{i=1}^2\!\frac{\partial}{\partial q_i}\!\left[\mu_i(u_i)\left(\Upsilon\tilde{\rho} \frac{\partial\mathcal{H}}{\partial q_i}+ \frac{\partial\Upsilon\tilde{\rho}}{\partial q_i}\right)\right]\!,\quad \label{eqa4}
\end{eqnarray}
plus higher order terms. The solvability condition for  the linear Eq.~\eqref{eqa4} is that the integral of its right hand side with respect to $x,p$ be zero. Using Eqs.~\eqref{eqa3} and after some algebra, this yields:
\begin{subequations}\label{eqa5}
\begin{eqnarray}
&&\frac{\partial\tilde{\rho}}{\partial t}=\nu\sum_{i=1}^2\!\frac{\partial}{\partial q_i}\!\left[\langle\mu_i(u_i)\rangle_x\!\left(  \frac{\partial\tilde{\rho}}{\partial q_i}-\frac{\left\langle\mu_i(u_i)u_i\right\rangle_x}{\langle\mu_i(u_i)\rangle_x}\tilde{\rho}\right)\right]\!, \quad\quad  \label{eqa5a}\\
&&\langle g(x)\rangle_x
= \frac{\int g(x)\, \exp\!\left[-U(x)-\frac{q^2(x+\lambda)-\Gamma(v)x}{2\lambda\Gamma(v)}\right] dx}{\int\exp\!\left[-U(x)-\frac{q^2(x+\lambda)-\Gamma(v)x}{2\lambda\Gamma(v)}\right] dx}. \label{eqa5b}
\end{eqnarray}\end{subequations}
The Ito SDEs corresponding to the averaged FPE \eqref{eqa5a} are
\begin{subequations}\label{eqa6}
\begin{eqnarray}
dq_i\!&=&\!\! \left[ \frac{\partial\langle\mu_i(u_i)\rangle_x}{\partial q_i} +\langle\mu_i(u_i)u_i\rangle_x\right]\! d(\nu t)\nonumber\\
&\!+&\!\sqrt{2\langle\mu_i(u_i)\rangle_x}\,dw_{q_i}(\nu t), \label{eqa6a}\\
u_i\!&=&\!-\left[\frac{q_1+q_2}{\Gamma(v)}\!\left(1+\frac{x}{\lambda}\right)\!+\nu q_i+v\right]\!. \label{eqa6b}
\end{eqnarray}
\end{subequations}
Eqs.~\eqref{eqa6} are circuit equations in which the mobilities and currents are replaced by their stochastic averages for the local equilibrium of the graphene variables given by Eq.~\eqref{eqa5b}. These equations agree with the stochastic averaging theorem \cite{kha64,bon14} and the numerical observation that the graphene is in local equilibrium with the instantaneous values of the charges in the circuit. In the limit $l\ll d$, $\lambda\to\infty$ and the averages $\langle\mu_i(u_i)\rangle$ and $\langle\mu_i(u_i)u_i\rangle$ coincide with $\mu_i(u_i)$ and $\mu_i(u_i)u_i$, respectively. Thus in this limit, we can replace $C(x)\approx C_0$ in the FPE \eqref{eqa1a}.

\section{One diode in the limit as $k_BT\ll Cu_0^2$: Propagation of equilibrium front}\label{ap_b}
The substitution
\begin{eqnarray}
\rho(q,t)= g(q,t) e^{-q^2/(2C_0k_BT)}, \label{eqb1}
\end{eqnarray}
transforms Eq.~\eqref{eq6} into 
\begin{eqnarray}
\frac{\partial g}{\partial t}+\frac{q}{C_0}\mu_0 \frac{\partial g}{\partial q}-k_BT \frac{\partial}{\partial q}\!\left(\mu_0\frac{\partial g}{\partial q}\right)\!=0. \label{eqb2}
\end{eqnarray}
Provided $k_BT\ll C_0u_0^2$ in Eqs.~\eqref{eqb1}, the last term on the right side of this equation can be ignored and we obtain the solution 
\begin{subequations}\label{eqb3}
\begin{eqnarray}
&&g\propto \Theta(Q(t)-q), \quad\mbox{  where } \label{eqb3a}\\
&& \dot{Q}=\frac{Q}{C_0}\mu_0=\frac{1}{RC_0}\,\frac{Q}{1+e^{Q/(C_0u_0)}}. \label{eqb3b}
\end{eqnarray}
The velocity of characteristics is exponentially small as $Q \to\infty$, so the front slows down dramatically as it advances. The time it takes the front to advance from $q = Q$ to $q = Q +\delta Q$, $0 <\delta Q\ll Q$, is approximated by
\begin{eqnarray}
\delta t \sim RC\frac{1+e^{Q/(C_0u_0)}}{Q} \, \delta Q. \label{eqb3c}
\end{eqnarray}
\end{subequations}
This time becomes exponentially large as $Q\to \infty$. The decay of the ensemble-averaged charge slows down due to the slowing of the front. If we approximate
\begin{eqnarray}
&&\rho= \frac{\Theta(Q(t)-q)}{\int_{-\infty}^Q e^{-q^2/(2C_0k_BT)}dq}\, e^{-q^2/(2C_0k_BT)}, \label{eqb4}
\end{eqnarray}
then the uniform value of $g$ behind the front is not exactly time-independent, consistent with the advection equation Eq.~\eqref{eqb2}. This is an error associated with the step function approximation to $g$. The error in normalization is exponentially small for $Q \gg \sqrt{C_0k_BT}$, and has negligible effect on the estimate of ensemble-averaged charge, which is now
\begin{eqnarray}
\langle q\rangle\!=\! \frac{\int_{-\infty}^Q\! q\, e^{-\frac{q^2}{2C_0k_BT}}dq}{\sqrt{2\pi C_0k_BT}}\!  =\!- \sqrt{\frac{C_0k_BT}{2\pi}} e^{-\frac{Q^2}{2C_0k_BT}}. \label{eqb5}
\end{eqnarray}
The relative increase of average charge over the time in which the front
advances from $q = Q$ to $q = Q + \delta Q$ is
\begin{subequations}\label{eqb6}
\begin{eqnarray}
\frac{\delta\langle q\rangle}{\langle q\rangle} \sim \frac{Q\,\delta Q}{C_0k_BT}. \label{eqb6a}
\end{eqnarray}
From Eqs.~\eqref{eqb3c} and \eqref{eqb6}, the time required for a given relative increase of charge, $a=\delta\langle q\rangle/\langle q\rangle$, is
\begin{eqnarray}
\delta t \sim RC_0^2k_BT \frac{1+e^{Q/(C_0u_0)}}{Q^2}\, a. \label{eqb6b}
\end{eqnarray}\end{subequations}
This time becomes exponentially large as $Q\to\infty$.

The diffusion in Eq.~\eqref{eqb2} smooths out the front at $q = Q(t)$. To see how this
works, we examine the equation for the gradient of $g$ near the front, $r=\partial g/\partial q$. Eq.~\eqref{eqb2} becomes
\begin{eqnarray}
\frac{\partial r}{\partial t}+\frac{\partial}{\partial q}\left(q\mu_0 r -k_BT\frac{\partial(\mu_0 r)}{\partial q}\right)\!=0. \label{eqb7}
\end{eqnarray}
This is a diffusion-convection equation (different from the FPE) and therefore $r$ is locally conserved. Hence, the total change in $g$ across the front is conserved. We represent $r$ in ``traveling wave'' form,
\begin{equation}
r=\mathcal{R}(\zeta,t), \quad \zeta=\frac{q-Q(t)}{\sqrt{C_0k_BT}}. \label{eqb8}
\end{equation}
$\mathcal{R}$ satisfies
\begin{widetext}
\begin{eqnarray*}
\frac{\partial\mathcal{R}}{\partial t}+\frac{1}{\sqrt{C_0k_BT}}\frac{\partial}{\partial\zeta}\!\left\{ \left[ \frac{Q+\sqrt{C_0k_BT}\zeta}{C}\mu\!\left(-\frac{Q+\sqrt{C_0k_BT}\,\zeta}{C_0}\right)\!-\dot{Q}\right]\!\mathcal{R} -\sqrt{\frac{k_BT}{C_0}}\frac{\partial}{\partial\zeta}\!\left[\mu\!\left(-\frac{Q+\sqrt{C_0k_BT}\,\zeta}{C_0}\right)\mathcal{R}\right]\right\}\!=0,
\end{eqnarray*}
or, evoking Eq.~\eqref{eqb3b} for $\dot{Q}$,
\begin{eqnarray}
\frac{\partial\mathcal{R}}{\partial t}+\frac{\partial}{\partial\zeta}\!\left\{\!\left[\frac{\frac{Q+\sqrt{C_0k_BT}\zeta}{C_0}\mu\!\left(-\frac{Q+\sqrt{C_0 k_B T}\,\zeta}{C_0}\right)\!-\frac{Q}{C_0}\mu\!\left(-\frac{Q}{C_0}\right)}{\sqrt{C_0k_BT}}\right]\!\mathcal{R} - \frac{\partial}{\partial\zeta}\!\left[\mu\!\left(-\frac{Q+\sqrt{C_0k_BT}\,\zeta}{C_0}\right)\!\frac{\mathcal{R}}{C_0}\right]\right\}\!=0. \quad   \label{eqb9}
\end{eqnarray}
\end{widetext}
In the limit as $C_0k_BT\to 0$,
\begin{eqnarray}
C_0\frac{\partial\mathcal{R}}{\partial t}- \frac{\partial}{\partial\zeta}\!\left\{(u\mu)'(u) \zeta\mathcal{R} + \frac{\partial}{\partial\zeta}[ \mu(u)\,\mathcal{R}]\right\}\!=0,  \label{eqb10}
\end{eqnarray}
where $u=-Q/C_0$ and $\mu(u)$ is given by Eq.~\eqref{eq6}. This equation has a Gaussian as solution, 
\begin{eqnarray}
\mathcal{R}=\frac{1}{\sqrt{2\pi\sigma}}\, e^{-\zeta^2/(2\sigma)},\label{eqb11}
\end{eqnarray}
 whose variance satisfies
\begin{eqnarray}
C_0\dot{\sigma}-2(u\mu)'(u)\,\sigma = 2\mu(u). \label{eqb12}
\end{eqnarray}
From Eq.~\eqref{eqb3b}, this equation becomes
\begin{eqnarray}
\frac{d\sigma}{dQ}-\frac{2(Q\mu)'(Q)}{Q\mu(Q)}\,\sigma = \frac{2}{Q}. \label{eqb13}
\end{eqnarray}
Suppose the front has advanced far enough so $Q\gg C_0u_0$, in which case $\mu(Q)\sim e^{-Q/(C_0u_0)}$, and Eq.~\eqref{eqb13} reduces to
\begin{eqnarray}
\frac{d\sigma}{dQ}+\!\left(\frac{2}{C_0u_0}-\frac{2}{Q}\right)\!\sigma = \frac{2}{Q}. \label{eqb14}
\end{eqnarray}
As $Q\to\infty$, there is an asymptotic solution 
\begin{eqnarray}
\sigma\sim \frac{C_0u_0}{Q}+\frac{3C_0^2u_0^2}{2Q^2}+\frac{3C^3_0u_0^3}{Q^3}+\ldots,  \label{eqb15}
\end{eqnarray}
and the front thickness narrows as it propagates further to the right.

\section{Scalings for the initial time stage and for longer times}\label{sec_c}
We now change variables to scaled charge sums and differences in Eq.~\eqref{eq5} for the complete circuit in Fig.~\ref{fig1} according to the definitions: 
\begin{subequations}\label{eqc1}
\begin{eqnarray}
&&\chi=\frac{q_1-q_2}{C_0V_0}, \quad\eta=\frac{q_1+q_2}{C_0V_0}(1+\epsilon),\quad s=\frac{t}{RC_0},\quad\label{eqc1a}\\
&&\epsilon=\frac{C_0}{2C}, \quad  V_0=\sqrt{\frac{k_BT}{C_0}}.\label{eqc1b}
\end{eqnarray}\end{subequations}
This scaling is appropriate for an initial time stage or layer provided $\epsilon\ll 1$. Eq.~\eqref{eq5} becomes
\begin{widetext}
\begin{subequations}\label{eqc2}
\begin{eqnarray}
&&\frac{\partial\rho}{\partial s}=\mathcal{L}_0\rho+ \epsilon\mathcal{N}_1\rho+\epsilon^2\mathcal{N}_2\rho,\label{eqc2a}\\
&&\mathcal{L}_0\!=\frac{\partial}{\partial\eta}\!\left[(\mu_1\!+\!\mu_2)\!\left( \frac{\partial}{\partial \eta}+ \eta\right)\!+(\mu_1\!-\!\mu_2)\frac{\partial}{\partial\chi} \right]\!+\frac{\partial}{\partial\chi}\!\left[(\mu_1\!-\!\mu_2)\!\left( \frac{\partial}{\partial \eta}+ \eta\right)\!+(\mu_1\!+\!\mu_2) \frac{\partial}{\partial\chi} \right]\!,\quad \quad\label{eqc2b}\\
&&\mathcal{N}_1\!= \frac{\partial}{\partial \eta}\!\left[(\mu_1\!-\!\mu_2)\!\left( \frac{\partial}{\partial\chi}\!+\!\chi\right)\!+(\mu_1\!+\!\mu_2)\!\left( 2\frac{\partial}{\partial \eta}+\eta\right)\!\right]\! \nonumber\\
&&\quad\,+ \frac{\partial}{\partial\chi}\!\left[(\mu_1\!-\!\mu_2)\!\left(2 \frac{\partial}{\partial\eta}\!+\!\eta\right)\!+(\mu_1\!+\!\mu_2)\frac{\partial}{\partial\chi}\right]\!, \quad  \label{eqc2c}\\
&&\mathcal{N}_2=\frac{\partial}{\partial\eta}\!\left[(\mu_1+\mu_2) \frac{\partial}{\partial\eta}+(\mu_1-\mu_2)\chi\right]\!,      \label{eqc2d}\\
&&\mu_i= \mu(-\eta+(-1)^i\epsilon\chi). \label{eqc2e}
\end{eqnarray}\end{subequations}
                \end{widetext}  
                For longer times, after the initial time stage, we have to change $\chi$ and the time $s$ to                       
\begin{equation}
\xi=\epsilon\chi=\frac{q_1-q_2}{2CV_0}, \quad\tilde{t}=\epsilon s= \frac{t}{2RC}.\label{eqc3}
\end{equation}
Then Eq.~\eqref{eq5} becomes
\begin{widetext}
\begin{subequations}\label{eqc4}
\begin{eqnarray}
&&\epsilon\frac{\partial\rho}{\partial t}=\mathcal{L}\rho+ \epsilon(2\mathcal{L}+\mathcal{M}_1)\rho+\epsilon^2(\mathcal{L}+\mathcal{M}_2) \rho,   \label{eqc4a}\\
&& \mathcal{L} = \frac{\partial}{\partial\eta} \!\left[(\mu_1+\mu_2)\!\left( \frac{\partial}{\partial \eta}+\eta\right)\! +(\mu_1-\mu_2)\xi\right]\!\label{eqc4b}\\
&&\mathcal{M}_1= \frac{\partial}{\partial\eta}\!\left[ (\mu_1-\mu_2)\!\left(\frac{\partial}{\partial\xi}-\xi\right)\!-(\mu_1+\mu_2)\eta\right]\! +\frac{\partial}{\partial\xi}\!\left[(\mu_1-\mu_2)\!\left(\frac{\partial}{\partial\eta}+\eta\right)\!+ (\mu_1+\mu_2)\xi\right]\!,\quad \label{eqc4c} \\
&&\mathcal{M}_2=  \frac{\partial}{\partial\eta} \!\left[(\mu_1-\mu_2) \!\left(\frac{\partial}{\partial\xi}-\xi\right)\! -(\mu_1+\mu_2)\eta \right]\!+\frac{\partial}{\partial\xi}\!\left[ (\mu_1+\mu_2)\frac{\partial}{\partial\xi}+(\mu_1-\mu_2)\frac{\partial}{\partial\eta}\right]\!,  \label{eqc4d}\\
&&\mu(u)=\frac{1}{1+e^{-u/w}},\quad w=\frac{u_0}{V_0},\quad \frac{u_j}{V_0}= (-1)^j\xi-\eta,\,\, j=1,2, \label{eqc4e}\\
&&\frac{\mathcal{H}}{k_BT}= \frac{\eta^2}{2(1+\epsilon)}+\frac{\xi^2}{2\epsilon},\quad \rho_\text{eq}(\eta,\xi)=\frac{1}{2\pi\sqrt{\epsilon(1+\epsilon)}} \exp\!\left[-\!\left(\frac{\eta^2}{2(1+\epsilon)}+\frac{\xi^2}{2\epsilon}\right)\right]\!.        \label{eqc4f}
\end{eqnarray}
\end{subequations}\end{widetext}
in which we have dropped the tilde in $\tilde{t}$. It is straightforward to check that $\rho_\text{eq}$ in Eq.~\eqref{eqc4f} is a stationary solution of Eq.~\eqref{eqc4a}. However, as $\epsilon\to 0$ and for diodes conducting in opposite directions, we shall find in Appendix \ref{ap_e} an approximate stationary solution that is different from equilibrium to the order in $\epsilon$ we consider.

\section{Initial layer} \label{ap_d}
For diodes conducting in opposite directions, the arguments of the mobility functions in Eq.~\eqref{eqc2e} are $\pm\eta-\epsilon\chi$, 
\begin{eqnarray}
\mu_1+\mu_2= 1+ O(\epsilon),\,\, \mu_1-\mu_2=-\tanh\frac{\eta}{2w}+O(\epsilon).\quad \label{eqd1}
\end{eqnarray}
\begin{widetext}
Then the leading order of Eq.~\eqref{eqc2a} is 
\begin{subequations}\label{eqd2}
\begin{eqnarray}
\frac{\partial\rho^{(0)}}{\partial s}=\frac{\partial}{\partial\eta}\!\left[\left( \frac{\partial}{\partial \eta}+ \eta\right)\!-\tanh\frac{\eta}{2w}\frac{\partial}{\partial\chi} \right]\!\rho^{(0)}-\frac{\partial}{\partial\chi}\!\left[\tanh\frac{\eta}{2w}\!\left(\frac{\partial}{\partial \eta}+\eta\right)\! - \frac{\partial}{\partial\chi}\right]\!\rho^{(0)}.\label{eqd2a}
\end{eqnarray}
We now substitute $\rho^{(0)}=e^{-\eta^2/2}R^{(0)}(\chi,\eta,s)$ in this equation and integrate the result with respect to $\eta$ to obtain an equation for the $\chi$-dependent reduced probability density. We find
\begin{eqnarray}
&&\int_{-\infty}^\infty e^{-\frac{\eta^2}{2}}\frac{\partial R^{(0)}}{\partial s}\,\frac{d\eta}{\sqrt{2\pi}}=\int_{-\infty}^\infty\!\left\{\frac{\partial}{\partial\eta}\!\left[e^{-\frac{\eta^2}{2}}\left( \frac{\partial R^{(0)}}{\partial \eta}- \tanh\frac{\eta}{2w}\frac{\partial R^{(0)}}{\partial\chi}\right)\!\right]\right.\nonumber\\ \quad\quad\nonumber\\
&&\quad+\left.e^{-\frac{\eta^2}{2}}\frac{\partial}{\partial\chi}\!\left[-\tanh\frac{\eta}{2w}\!\left(\frac{\partial R^{(0)}}{\partial \eta}-\tanh\frac{\eta}{2w}\frac{\partial R^{(0)}}{\partial\chi}\right)\!+\mbox{sech}^2\frac{\eta}{2w}\frac{\partial R^{(0)}}{\partial\chi}\right]\!\right\} \frac{d\eta}{\sqrt{2\pi}}.\label{eqd2b}
\end{eqnarray}
\end{subequations}
\end{widetext}
Assuming that $R^{(0)}$ is a function of $s$ and of the new variable $\sigma=\chi+2w\ln\cosh\frac{\eta}{2w}$, all terms on the right hand side of Eq.~\eqref{eqd2b} cancel except for the last one. Then we obtain the heat equation:
\begin{subequations}
\begin{eqnarray}
&&\frac{\partial R^{(0)}}{\partial s}=a\,\frac{\partial^2 R^{(0)}}{\partial\sigma^2}, \label{eqd3a}\\
&&\sigma=\chi+2w\ln\cosh\frac{\eta}{2w},\label{eqc2b}\\
&& a=\frac{1}{\sqrt{2\pi}} \int_{-\infty}^\infty e^{-\frac{\eta^2}{2}}\mbox{sech}^2\frac{\eta}{2w} d\eta.\label{eqd3c}
\end{eqnarray}\end{subequations}
For a delta-function initial condition corresponding to the initial zero charge in the circuit, the Gaussian kernel solves this equation and produces the normalized solution
\begin{eqnarray}
\rho^{(0)}(\chi,\eta,s)= \frac{e^{-\frac{\eta^2}{2}}}{2\pi\sqrt{2 a s}} \exp\!\left[-\frac{\left(\chi\!+\!2w\ln\cosh\frac{\eta}{2w}\right)^2}{4as}\right]\!. \quad\quad \label{eqd4}
\end{eqnarray}
This probability density yields the averages:
\begin{subequations}\label{eqd5}
\begin{eqnarray}
&&\langle\eta\rangle=0,\quad\langle\sigma^2\rangle=2a s,\label{eqd5a}\\
&&\langle\chi\rangle=-\frac{2w}{\sqrt{2\pi}} \int_{-\infty}^\infty e^{-\frac{\eta^2}{2}}\ln\cosh\frac{\eta}{2w} d\eta\underbrace{\sim}_{w\to 0} -\sqrt{\frac{2}{\pi}}.\quad\quad\label{eqd5b}
\end{eqnarray}\end{subequations}
According to Eq.~\eqref{eqd5a}, the average charge at the capacitor $C_0$ is zero, capacitor 1 has negative average charge and capacitor 2 has positive average charge of the same magnitude (the opposite signs to charges due to a battery if we ignore thermal fluctuations). This is a surprising result: the system does not evolve to the equilibrium $e^{-\eta^2/2}/\sqrt{2\pi}$. Instead, this initial layer builds up opposite charges at the capacitors and the variance of the state \eqref{eqd5a} increases linearly with time. Note that, in dimensional units, the charge \eqref{eqd5b} yields $\langle q_1\rangle\sim-(1/2)\sqrt{2k_BTC_0/\pi}$, which is half the charge for the case of a single perfectly conducting diode (with piecewise linear current-voltage curve), except that the capacitor in series with the diode has been replaced by the small capacitance $C_0$.

We can calculate the average energy rate and the entropy production from Eqs.~\eqref{eqd2a} and \eqref{eqd4} using integration by parts. The results are:
\begin{eqnarray}
&&\frac{d}{ds}\langle\mathcal{H}\rangle= O(\epsilon),\nonumber\\
&& \frac{dS}{ds}=-\frac{d}{ds}\int \rho\,\ln\rho\, d\eta\,d\chi=\frac{1}{as} + O(\epsilon).
\label{eqd6} \end{eqnarray}
The production of entropy declines as time elapses. Thus, the entropy increases to a large value after $t=0$ and then it increases  logarithmically as $s\to\infty$ at the end of the initial stage. Direct numerical simulations of the stochastic equations show an initial build-up of entropy followed by stabilization in Fig.\! 4(d).

In the long time scaling \eqref{eqc3} with the variables $\xi$ and $t$, Eq.~\eqref{eqd4} becomes
\begin{subequations}\label{eqd7}
\begin{eqnarray}
\rho^{(0)}(\xi,\eta,t)\!&=&\! \frac{e^{-\frac{\eta^2}{2}}}{2\pi\sqrt{2 a\epsilon t}}\,\exp\!\left[-\frac{\left(\xi+2w\epsilon\ln\cosh\frac{\eta}{2w}\right)^2}{4a\epsilon t}\right]\nonumber\\
&\sim&\! \frac{1}{\sqrt{2\pi}}e^{-\frac{\eta^2}{2}}\delta(\xi), \label{eqd7a}
\end{eqnarray}
as $\epsilon \to 0$ and $t=O(1)$. Eqs.~\eqref{eqd5} become
\begin{eqnarray}
&&\langle\eta\rangle=0,\quad \langle\xi\rangle=-\frac{2\epsilon w}{\sqrt{2\pi}}\! \int_{-\infty}^\infty\! e^{-\frac{\eta^2}{2}}\!\ln\cosh\!\frac{\eta}{2w} d\eta, \label{eqd7b}\\
&&\langle(\xi-\langle\xi\rangle)^2\rangle=\frac{2\epsilon t}{\sqrt{2\pi}}\! \int_{-\infty}^\infty\! e^{-\frac{\eta^2}{2}}\mbox{sech}^2\! \frac{\eta}{2w}\, d\eta\nonumber\\
&&\quad\quad\quad\quad\quad+\frac{4\epsilon^2 w^2}{\sqrt{2\pi}}\! \left[\int_{-\infty}^\infty\! e^{-\frac{\eta^2}{2}}\!\left(\ln\cosh\!\frac{\eta}{2w}\right)^2 d\eta\right.\nonumber\\
&&\quad\quad\quad\quad\quad -\left.\frac{1}{\sqrt{2\pi}}\!\left( \int_{-\infty}^\infty\! e^{-\frac{\eta^2}{2}}\!\ln\cosh\!\frac{\eta}{2w} d\eta\right)^2\right]\!.\quad \label{eqd7c}
\end{eqnarray}
\end{subequations}
                                                                                                                                                                                                                                                                                                                                                                                                                                                                                                                                                                                                                                                                                                                                                                                                                                                                                                                                                                                                                                                                                                                                                                        
\section{Long time scaling and quasistationary probability density} \label{ap_e}
After the initial layer described in Appendix \ref{ap_d}, and for appropriate small values of $\epsilon$ and $w$, the numerical solution of the FPE indicates that the probability density produces a very flat maximum of the average capacitor charge before decreasing to zero (thermal equilibrium); see Fig.~\ref{fig2}. For sufficiently small values of $\epsilon$ and $w$, the flat maximum corresponds to a quasi-stationary solution of the FPE, which we derive in this section.

If the diodes are conducting in opposite directions, we use the scaling \eqref{eqc3} producing Eqs.~\eqref{eqc4}. The stationary solution of $\mathcal{L}\rho=0$ is 
\begin{eqnarray}
E(\xi,\eta)\!&=&\! e^{-\frac{\eta^2}{2}}\exp\!\left[-\xi \int_0^\eta \frac{\mu(-s-\xi)-\mu(s-\xi)}{\mu(-s-\xi)+\mu(s-\xi)}\,ds \right] \!\nonumber\\
\!&=& e^{-\frac{\eta^2}{2}}\!\left(\frac{e^{-\frac{\xi}{w}}+\cosh\frac{\eta}{w}}{1+e^{-\frac{\xi}{w}}}\right)^{\xi w}\!, \label{eqe1}
\end{eqnarray}
which is integrable in the variable $\eta$. Moreover, the solution of the leading univariate FPE in fast time scale $t/\epsilon$, $\partial\rho/\partial (t/\epsilon)=\mathcal{L}\rho$, tends to a normalized version of \eqref{eqe1} as $t/\epsilon\to \infty$:
\begin{subequations} \label{eqe2}
\begin{eqnarray}
&&\rho^{(0)}=\hat{E}(\xi,\eta)\, P(\xi,t;\epsilon)=\frac{E(\xi,\eta)P(\xi,t;\epsilon)}{\int_{-\infty}^\infty E(\xi,\eta) d\eta}, \label{eqe2a}\\
&&\int_{-\infty}^\infty P(\xi,t;\epsilon)\, d\xi =1. \label{eqe2b}
\end{eqnarray}\end{subequations}
Clearly, the average of $\eta$ using Eq.~\eqref{eqe2a} is zero because $E(\xi,\eta)$ is even in $\eta$. So the charge at the capacitor $C_0$ is zero. 

\subsection{Reduced FPE by the Chapman-Enskog method}
To find the reduced equation for the slowly varying probability density $P(\xi;\epsilon)$, we use the Chapman-Enskog method:
\begin{subequations} \label{eqe3}
\begin{eqnarray}
&&\rho= \rho^{(0)}(\xi,\eta;P)+ \sum_{j=1}^2\epsilon^j\rho^{(j)}(\eta;P)+O(\epsilon^3),  \label{eqe3a}\\
&& \frac{\partial P}{\partial t}= F^{(0)}+ \epsilon F^{(1)}+O(\epsilon^2), \label{eqe3b}\\
&&\int_{-\infty}^\infty \rho^{(j)}(\eta;P)\, d\eta=0, \label{eqe3c}
\end{eqnarray}\end{subequations}
where the $F^{(j)}$ are functionals of $P$ selected so that the hierarchy of linear equations for the $\rho^{(j)}$ have bounded solutions. The result is  
\begin{subequations}\label{eqe4}
\begin{eqnarray}
&&\frac{\partial P}{\partial t}=\frac{\partial}{\partial\xi}\!\left[M(\xi;\epsilon)\, P+\epsilon D(\xi)\,\frac{\partial P}{\partial\xi}\right]\!, \label{eqe4a}\\
&& \int_{-\infty}^\infty P(\xi,t;\epsilon)\, d\xi =1. \nonumber
\end{eqnarray}
The stationary solution of Eq.~\eqref{eqe4a} is
\begin{eqnarray}
P_s(\xi;\epsilon)= \frac{1}{Z}\,\exp\!\left(-\int \frac{M(\xi;\epsilon)}{\epsilon\, D(\xi)}\, d\xi\right)\!.   \label{eqe4b}
\end{eqnarray}
This probability density is a globally stable solution of the reduced equation \eqref{eqe4a} because the following relative entropy is a Lyapunov functional:
\begin{eqnarray}
\mathcal{F}[P]\!&=&\!\int_{-\infty}^\infty\!\int_{-\infty}^\infty\!\hat{E}P\ln\!\left(\frac{\hat{E}P}{\hat{E}P_s}\right)d\eta\, d\xi\nonumber\\
\!&=&\!\int_{-\infty}^\infty P(\xi,t;\epsilon)\ln\!\left(\frac{P(\xi,t;\epsilon)}{P_s(\xi;\epsilon)}\right) d\xi. \label{eqe4c}
\end{eqnarray}
\end{subequations}
Thus, the reduced probability density evolves towards $P_s(\xi;\epsilon)$ from the initial condition $P_s(0;\epsilon)=\delta(\xi)$, which is compatible with the initial layer of Eq.~\eqref{eqd7a}.

The drift and diffusion coefficients in the reduced FPE \eqref{eqe4a} are 
\begin{widetext}
\begin{subequations} \label{eqe5}
\begin{eqnarray}
&&M(\xi;\epsilon)=4\xi\!\int_{-\infty}^\infty\! \frac{\mu_1\mu_2 \hat{E}}{\mu_1+\mu_2}d\eta+\epsilon \int_{-\infty}^\infty\!\frac{\hat{E}}{\mu_1+\mu_2}\!\left\{ 4(\mu_1-\mu_2)\frac{\Psi}{\hat{E}}+4 \mu_1\mu_2  \frac{\partial\ln\hat{E}}{\partial\xi}\right.\nonumber\\
&& \quad\quad\quad\quad +(\mu_1^2-\mu_2^2)\eta + (\mu_1-\mu_2)^2\xi-(\mu_1+\mu_2)\frac{\partial(\mu_1-\mu_2)}{\partial\eta}+4\xi\mu_1\mu_2 \nonumber\\
&& \quad\quad\quad\quad \left.\times \!\left[\frac{\eta^2}{2}-\! \int_{-\infty}^\infty \frac{\eta^2}{2} \hat{E}d\eta+\!\int_0^\eta\!\frac{ \frac{4\Psi}{\hat{E}}+(\mu_1-\mu_2)\!\left(\xi-\frac{\partial\ln\hat{E}}{\partial\xi}\right)\!}{\mu_1+\mu_2}d\eta'\right. \right. \nonumber\\
&&\quad\quad\quad\quad\left. \left.-\int_{-\infty}^\infty\!d\eta\,\hat{E}\!\int_0^\eta\!\frac{ \frac{4\Psi}{\hat{E}}+(\mu_1-\mu_2)\!\left(\xi-\frac{\partial\ln\hat{E}}{\partial\xi}\right)\!}{\mu_1+\mu_2} \right]\!\right\}d\eta,
\label{eqe5a}\\
&&  \Psi(\xi,\eta)=\frac{\partial}{\partial\xi}\!\left(\xi\int_\eta^\infty \frac{\mu_1\mu_2 \hat{E}d\eta'}{\mu_1+\mu_2} \right)\!-\!\left(\int_\eta^\infty\hat{E}d\eta'\right)\frac{\partial}{\partial\xi}\!\left(\xi\int_{-\infty}^\infty \frac{\mu_1\mu_2 \hat{E}d\eta'}{\mu_1+\mu_2}\right)\!,             \label{eqe5b}
\end{eqnarray}
\begin{eqnarray}
&&D(\xi)= \int_{-\infty}^\infty\frac{4\mu_1\mu_2\hat{E}}{\mu_1+\mu_2}\!\left[1+\xi\!\left(\frac{1}{\mu_2}-\frac{1}{\mu_1}\right)\!\Phi+4\xi^2 \!\left(\int_0^\eta\!\frac{\Phi d\eta'}{(\mu_1+\mu_2)\hat{E}}\right.\right.\quad\nonumber\\
&&\quad\quad\quad\left.-\int_{-\infty}^\infty\!\hat{E}\!\int_0^\eta\!\frac{\Phi d\eta'}{(\mu_1+\mu_2)\hat{E}}\right)\!\left.- \xi\!\left(\int_0^\eta\!\frac{\mu_1-\mu_2}{\mu_1+\mu_2}d\eta'-\!\int_{-\infty}^\infty \!d\eta\hat{E}\!\int_0^\eta\frac{\mu_1-\mu_2 }{\mu_1+\mu_2}d\eta'\right)\right]\!d\eta, \quad    \label{eqe5c}\\
&&\Phi(\xi,\eta)=\int_\eta^\infty \frac{\mu_1\mu_2 \hat{E}d\eta'}{\mu_1+\mu_2} -\!\left(\int_{-\infty}^\infty\frac{\mu_1\mu_2 \hat{E}d\eta'}{\mu_1+\mu_2}\right)\int_\eta^\infty\hat{E}d\eta,   \quad \Phi(\xi,-\infty)=0.     \label{eqe5d}
\end{eqnarray}
If we have only one diode, say $\mu_2=0$, $M=D=0$ as the scaling \eqref{eqc3} does not make sense. Note that $(\mu'_1-\mu'_2)$ is large and therefore the dominant terms in the drift coefficient \eqref{eqe5a} are
\begin{eqnarray}
M(\xi;\epsilon)\approx 4\xi\!\int_{-\infty}^\infty\! \frac{\mu_1\mu_2 \hat{E}}{\mu_1+\mu_2}d\eta-\epsilon \int_{-\infty}^\infty \frac{\partial(\mu_1-\mu_2)}{\partial\eta}\,\hat{E}\, d\eta.  \label{eqe5e}
\end{eqnarray}
\end{subequations}

\begin{figure}[h]
\begin{center}
\includegraphics[width=7cm]{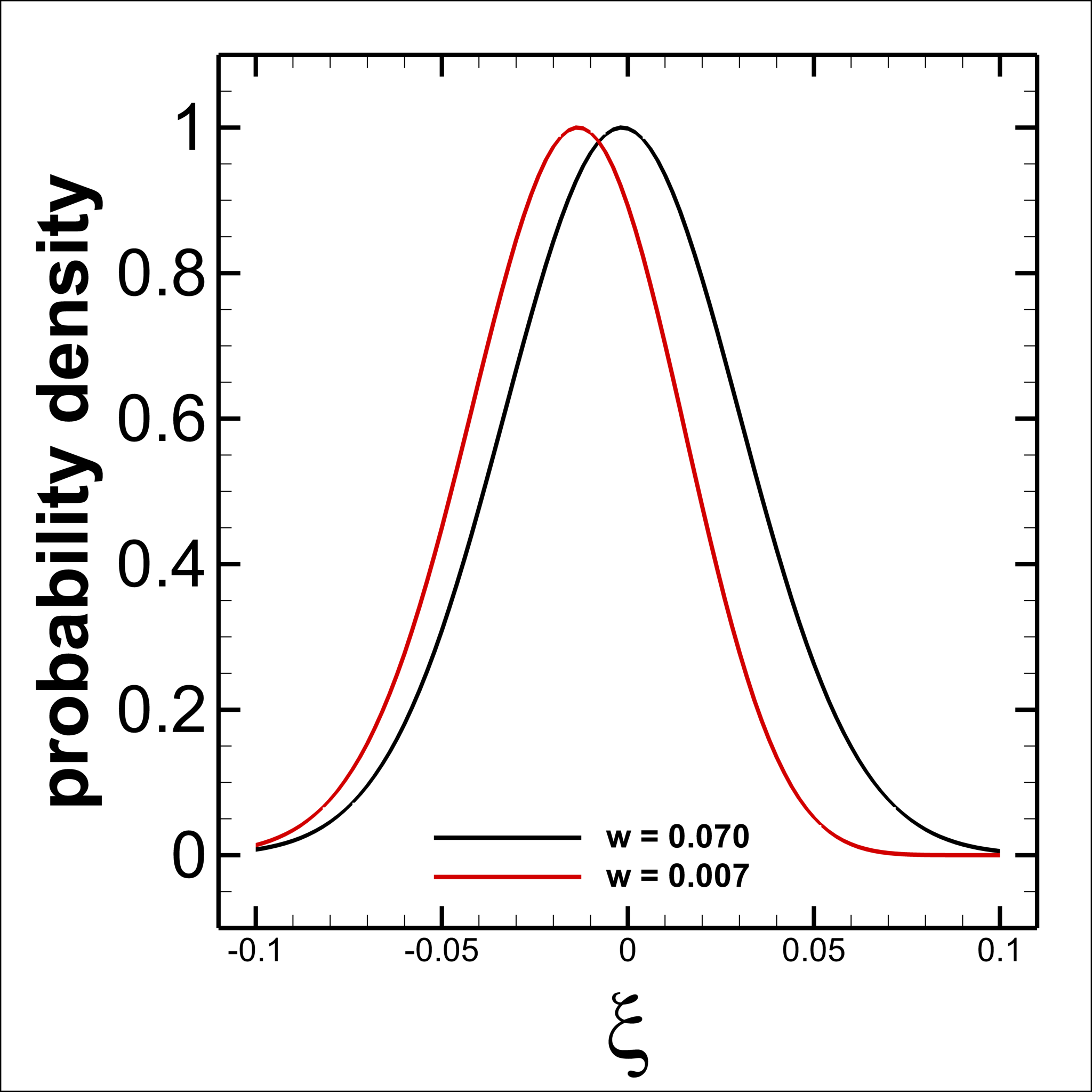}
\end{center}
\caption{Reduced stationary probability density $P_s(\xi)$ for two different values of $w$. Lowering $w$ shifts the maximum of $P_s(\xi)$ thereby yielding nonzero average charge $\langle\xi\rangle$. \label{fig5}}
\end{figure}
\end{widetext}

The reduced stationary probability density has to be calculated numerically because it cannot be approximated by a Gaussian function with a large second derivative about its maximum. Then the integrals entering the stationary averages over $\xi$ cannot be approximated by the usual expansion about the maximum of the integrand. Fig.~\ref{fig5} depicts the reduced stationary probability density $P_s(\xi)$ of Eq.~\eqref{eqe4b} for two different values of $w$. Lowering $w$ shifts the maximum of $P_s(\xi)$ to the left, which yields a negative average charge $\langle q_1\rangle= -C\sqrt{T/c}\, |\langle\xi\rangle|$ at the first capacitor and a positive charge $\langle q_2\rangle= C\sqrt{T/c}\, |\langle\xi\rangle|$ at the second capacitor. Notice that the signs of these charges coincide with those provided by the initial layer of Appendix \ref{ap_d} and are contrary to those of the charges produced by a noiseless circuit with a battery as in Fig.~\ref{fig1}(a).
\bigskip

\subsection{Calculation of the coefficient functions in the reduced FPE}
How do we find the coefficients in Eq.~\eqref{eqe4a}? Inserting Eqs.~\eqref{eqe3a} and \eqref{eqe3b} into \eqref{eqc4a}, we obtain Eq.~\eqref{eqe2a} and the hierarchy
\begin{subequations}\label{eqe6}
\begin{eqnarray}
&&\mathcal{L}\rho^{(1)}\!=F^{(0)}\hat{E}- (2\mathcal{L}+\mathcal{M}_1)\rho^{(0)},   \label{eqe6a}\\
&& \mathcal{L}\rho^{(2)}\! = F^{(1)}\!\hat{E} + \frac{\delta\rho^{(1)}}{\delta P} F^{(0)}\!-(2\mathcal{L}+\mathcal{M}_1)\rho^{(1)}\nonumber\\
&&\quad\quad- (\mathcal{L}+\mathcal{M}_2) \rho^{(0)}\!,        \label{eqe6b}
\end{eqnarray}
\end{subequations}
etc. The solvability condition for Eq.~\eqref{eqe6a} is that the integral of its right hand side over $\eta\in\mathbb{R}$ vanish. This yields
\begin{subequations}\label{eqe7}
\begin{eqnarray}
&&F^{(0)}= 4\frac{\partial}{\partial\xi}\!\left[\xi P\, \overline{\!\left(\frac{\mu_1\mu_2\hat{E}}{\mu_1+\mu_2} \right)\!}\right]\!, \label{eqe7a}\\
&&\overline{f(\eta)}=\int_{-\infty}^\infty f(\eta)\, d\eta. \label{eqe7b}
\end{eqnarray}
\end{subequations}

\begin{widetext}
The solution of Eq.~\eqref{eqe6a} that satisfies \eqref{eqe3c} is
\begin{subequations}\label{eqe8}
\begin{eqnarray}
\rho^{(1)}= \hat{E}\left(R-\overline{\hat{E} R}\right)\!,\label{eqe8a}
\end{eqnarray}
where 
\begin{eqnarray}
&&\frac{\partial R}{\partial\eta}= \frac{4\xi\frac{\Phi}{\hat{E}}-\mu_1+\mu_2}{\mu_1+\mu_2} \frac{\partial P}{\partial\xi} + \!\left[\eta+\frac{4\frac{\Psi}{\hat{E}}+(\mu_1-\mu_2)\!\left(\xi-\frac{\partial\ln\hat{E}}{\partial\xi}\right)\!}{\mu_1+\mu_2} \right] P,\label{eqe8b}\\
&& \rho^{(1)}= \left[\int_0^\eta \frac{4\xi\frac{\Phi}{\hat{E}}-\mu_1+\mu_2}{\mu_1+\mu_2}-\overline{\hat{E} \int_0^\eta \frac{4\xi\frac{\Phi}{\hat{E}}-\mu_1+\mu_2}{\mu_1+\mu_2} }\right]\hat{E}\frac{\partial P}{\partial\xi} \nonumber\\
&&\,\,+\!\left[\frac{\eta^2}{2}-\overline{\frac{\eta^2\hat{E}}{2}}+\! \int_0^\eta\frac{4\frac{\Psi}{\hat{E}}\! +\! (\mu_1\! -\! \mu_2)\!\left(\xi\! -\! \frac{\partial\ln\hat{E}}{\partial\xi}\right)\!}{\mu_1+\mu_2}-\overline{\hat{E}\! \int_0^\eta\frac{4\frac{\Psi}{\hat{E}}\! +\! (\mu_1\! -\! \mu_2)\!\left(\xi\! -\! \frac{\partial\ln\hat{E}}{\partial\xi}\right)}{\mu_1+\mu_2}}\,\right]\!\hat{E}P. \quad\label{eqe8c}
\end{eqnarray}
\end{subequations}

After integration by parts, the solvability condition for Eq.~\eqref{eqe6b} yields
\begin{eqnarray}
&&F^{(1)}= \frac{\partial}{\partial\xi}\!\left\{\overline{ (\mu_1+\mu_2)\hat{E}}\,\frac{\partial P}{\partial\xi}+\!\left[\, \overline{(\mu_1+\mu_2) \frac{\partial\hat{E}}{\partial\xi}-(\mu'_1-\mu'_2)\hat{E}}\,\right]P +\,\overline{\frac{4\mu_1\mu_2\xi}{\mu_1+\mu_2} \rho^{(1)}} 
+\,\overline{ (\mu_1-\mu_2)\hat{E} \frac{\partial R}{\partial\eta}}\,\right\}\!. \label{eqe9}
\end{eqnarray}\end{widetext}
We now insert Eqs.~\eqref{eqe7a} and \eqref{eqe9} in \eqref{eqe3b} and use Eqs.~\eqref{eqe8}, thereby obtaining the drift and diffusion coefficients of Eqs.~\eqref{eqe5} after some algebra. 

\section{Data from stochastic simulations and from experiments}\label{ap_f}
\begin{figure}[h]
\begin{center}
\includegraphics[width=7cm,angle=0]{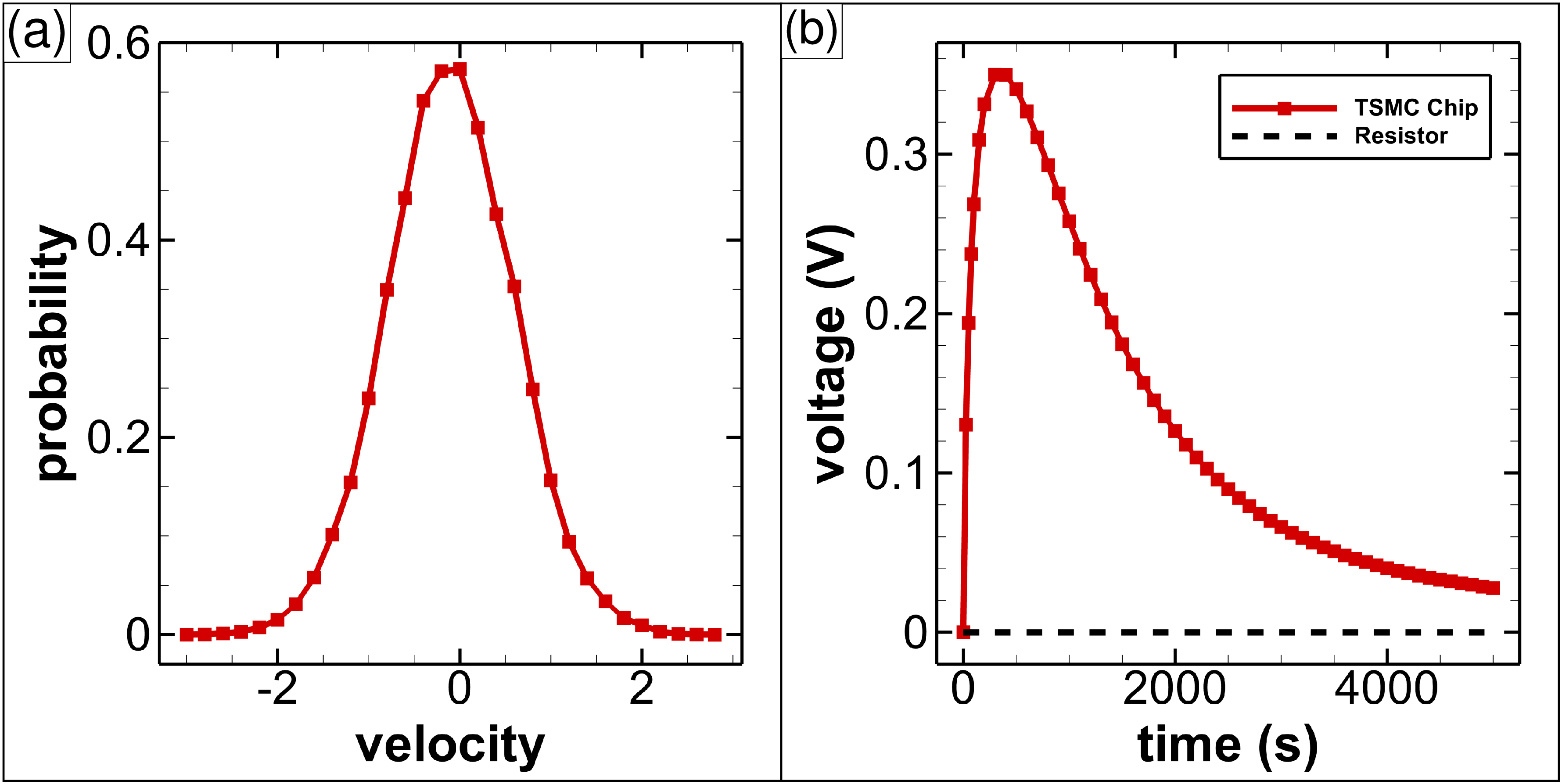}
\end{center}
\caption{(a) Measured histogram of the numerically simulated velocity distribution for graphene. (b) Measured voltage versus time plot for a chip comprising a single diode-capacitor circuit (as control, same plot for a resistor). \label{fig6}}
\end{figure}

Figure \ref{fig6}(a) shows the simulated velocity distribution of graphene after a short time (1\% of the total simulation time). It is a Gaussian corresponding to local equilibrium as explained in Appendix \ref{ap_a}. 

Figure \ref{fig6}(b) depicts the time evolution of the voltage across the capacitor (proportional to stored charge) measured using the single diode-capacitor circuit. We observe that it follows the theoretical curve for the inset of Fig.~\ref{fig1}(a).

\end{document}